\documentclass[preprint,amsmath,amssymb,prb,floatfix,superscriptaddress]{revtex4}

\usepackage[dvips]{graphicx}
\usepackage{dcolumn}
\usepackage{bm}
\usepackage{mathrsfs}

\begin{document}
\title{The role of spin fluctuations in the conductivity of CrO$_2$}
\author{Kate H. Heffernan}
 \affiliation{Department of Physics and Engineering Physics, Tulane University, 6400 Freret St., New Orleans, LA 70118, USA}
\author{Shukai Yu}
 \affiliation{Department of Physics and Engineering Physics, Tulane University, 6400 Freret St., New Orleans, LA 70118, USA}
\author{Skylar Deckoff-Jones}
 \affiliation{Department of Physics and Engineering Physics, Tulane University, 6400 Freret St., New Orleans, LA 70118, USA}
\author{Xueyu Zhang}
 \affiliation{Department of Chemistry, University of Alabama, Tuscaloosa, AL, USA}
\author{Arunava Gupta}
 \affiliation{Department of Chemistry, University of Alabama, Tuscaloosa, AL, USA}
\author{D. Talbayev}
 \email{dtalbayev@gmail.com}
 \affiliation{Department of Physics and Engineering Physics, Tulane University, 6400 Freret St., New Orleans, LA 70118, USA}

\date{\today}

\newcommand{\cm}{\:\mathrm{cm}^{-1}}
\newcommand{\T}{\:\mathrm{T}}
\newcommand{\mc}{\:\mu\mathrm{m}}
\newcommand{\ve}{\varepsilon}
\newcommand{\dg}{^\mathtt{o}}

\begin{abstract}
We present a time-resolved terahertz spectroscopic study of the half-metallic ferromagnet CrO$_2$.  The ultrafast conductivity dynamics excited by an optical pump displays a very short (several picoseconds) and a very long (several hundred picoseconds) characteristic time scales.  We attribute the former to the electron-phonon relaxation and the latter to the spin-lattice relaxation.  We use this distinction to quantify the relative contribution of the scattering by spin fluctuations to the resistivity of CrO$_2$: we find that they contribute less than one half of all scattering events below room temperature.  This contribution rises to $\sim70$\% as the temperature approaches $T_C$=390 K.  The small effect of spin fluctuations on the resistivity is unexpected in the light of the proposed double-exchange nature of the electronic and magnetic properties of CrO$_2$. 
\end{abstract}

\maketitle

\section{Introduction}
Chromium dioxide CrO$_2$ is a half-metallic ferromagnet ($T_C$=390 K), in which the majority spin electrons are metallic while the minority spin electrons are semiconducting, i.e., the Fermi level falls within a gap in the minority density of states\cite{schwarz:l211}.  The nearly 100\% spin polarization\cite{kamper:2788,guntherodt:247,soulen:85,anguelouch:180408,parker:196601} makes CrO$_2$ attractive as the source of spin-polarized electrons in spintronics, while the material was also used as the magnetic recording medium.  Theory predicted\cite{schwarz:l211} the magnetic moment per Cr$^{4+}$ ($3d^2$) ion to be 2$\mu_B$, in agreement with Hund's rules and experiment\cite{chamberland:1}.  Of the two $d$ electrons, one is localized and found about 1 eV below the Fermi level. The other $d$ electron hybridizes with the oxygen $p$ orbitals and forms a narrow itinerant band that crosses the Fermi level.  Korotin et al.\cite{korotin:4305} used the term "a self-doped double exchange ferromagnet" to describe the material's intertwined metallicity and ferromagnetism, with the mobile $d$ electrons mediating the double exchange between the localized $d$ spins.  Another remarkable feature is the two-order-of-magnitude drop in resistivity between 400 K and 10 K (Fig.~\ref{fig:drude}) whose origin is not fully understood.  In this work, we use time-resolved terahertz spectroscopy (TRTS) to compare the relative roles of spin fluctuation (or spin-flip) and other scattering processes in the resistivity of CrO$_2$.  We find that the spin-flip processes do not dominate the electron scattering in a wide range of temperatures below $T_C$, as many authors have assumed. 

CrO$_2$ crystallizes in the tetragonal rutile structure, with lattice parameters $a=b=0.4421$ nm and $c=0.2916$ nm\cite{anwar:085123}.  The Cr atoms are octahedrally coordinated by oxygen, and edge-sharing oxygen octahedra form ribbons along the $c$ axis, while the octahedra on adjacent ribbons share a corner\cite{lewis:10253}.  The Fermi level for the majority spins belongs in a half-filled band derived from the the $d_{xy}$ and $d_{yz}$ orbitals\cite{korotin:4305,lewis:10253}.  The band gap in the minority density of states exceeds 2 eV, with the empty minority states lying about 1 eV higher than the Fermi level\cite{korotin:4305,lewis:10253}.  The half-metallicity of CrO$_2$ was confirmed experimentally by point contact Andreev spectroscopy\cite{soulen:85}, tunneling measurements\cite{guntherodt:247}, and spin-polarized photoemission\cite{kamper:2788}.  

A survey of literature finds no agreement on the origin of the temperature dependence of resistivity (Fig.~\ref{fig:drude}).  Lewis et al.\cite{lewis:10253} showed that below about 200 K, the temperature dependence is well described by the Bloch-Gruneisen function and phonon scattering dominates.  Above 200 K, spin-flip scattering becomes important and contributes one half of all scattering events near the Curie temperature\cite{lewis:10253,coey:8345}.  Barry et al.\cite{barry:7166} suggested a phenomenological description based on the formula $\rho(T)=\rho_0+AT^2\exp(-\Delta/T)$ with a gap $\Delta\approx80$ K, above which the resistivity follows the $T^2$ dependence expected for a spin-flip scattering in a metallic ferromagnet. Gupta et al.\cite{gupta:6073} fit the low-temperature resistivity (below 40 K) with a $\rho(T)=\rho_0+AT^3$ dependence characteristic of spin-flip scattering if the non-rigid band behavior of the minority band is accounted for.  Watts et al. proposed a two-band picture for electronic conduction based on a magnetotransport study\cite{watts:9621}, although other magnetoresistance studies have not reached the same conclusion\cite{yanagihara:187201,branford:227201}.  Several authors found that a $T^2$ dependence also describes well the resistivity data in a broad temperature range and attributed this to electron-electron scattering\cite{suzuki:11597,anwar:085123}.

The contradictory scenarios proposed for electron conduction in CrO$_2$ perhaps reflect the reality that all three scattering processes - electron-electron, electron-phonon, and spin-flip - play a role.  Our TRTS study is motivated by the possibility of separating the different scattering contributions based on the different time scales for the coupling of electrons, lattice, and spins to the optical pump excitation.  When a metal is excited by the optical pump, the absorbed photons deposit their energy into the electronic system.  Within a picosecond, the relaxation of this energy establishes a thermal electron and phonon distribution at an elevated temperature\cite{talbayev:340,burch:026409,costa:104407}.  The subsequent thermalization of spins happens much slower, on the scale from tens to hundreds of picoseconds\cite{averitt:1357,averitt:017401}.  This vastly slower spin thermalization allows us to distinguish the contribution of the spin-flip scattering to resistivity from the contributions of the electron-electron and electron-phonon scattering.

\section{Experimental details and results}
For this study, we used an epitaxial 100 nm CrO$_2$ thin film grown on a (100)-oriented 0.5 mm TiO$_2$ substrate using chemical vapor deposition with CrO$_3$ as precursor\cite{gupta:6073}. TRTS and terahertz time-domain spectroscopy (THz TDS) measurements were performed using a home-built spectrometer based on an amplified Ti:sapphire laser with 1 kHz repetition rate\cite{silwal:092116}.  The THz wave was polarized along the crystalline $b$ axis of CrO$_2$. The optical pump pulses with 800 nm wavelength (1.55 eV photon energy) and 0.2 mJ/cm$^2$ fluence were polarized along the $c$ axis.  The THz probe spot diameter was 2 mm, while the diameter of the optical pump spot was 10 mm.  Temperature control in the 9-400 K range was provided by a He flow or a closed cycle cryostat.   

Time-resolved magneto-optical Kerr effect measurements (MOKE) were carried out at room temperature in the polar MOKE configuration.  The pump and probe wavelength was 800 nm.  The static applied magnetic field was normal to the film and measured to be $3300$ Gauss. It was supplied by a stack of permanent magnets.  

\begin{figure}[ht]
\begin{center}
\includegraphics[width=6in]{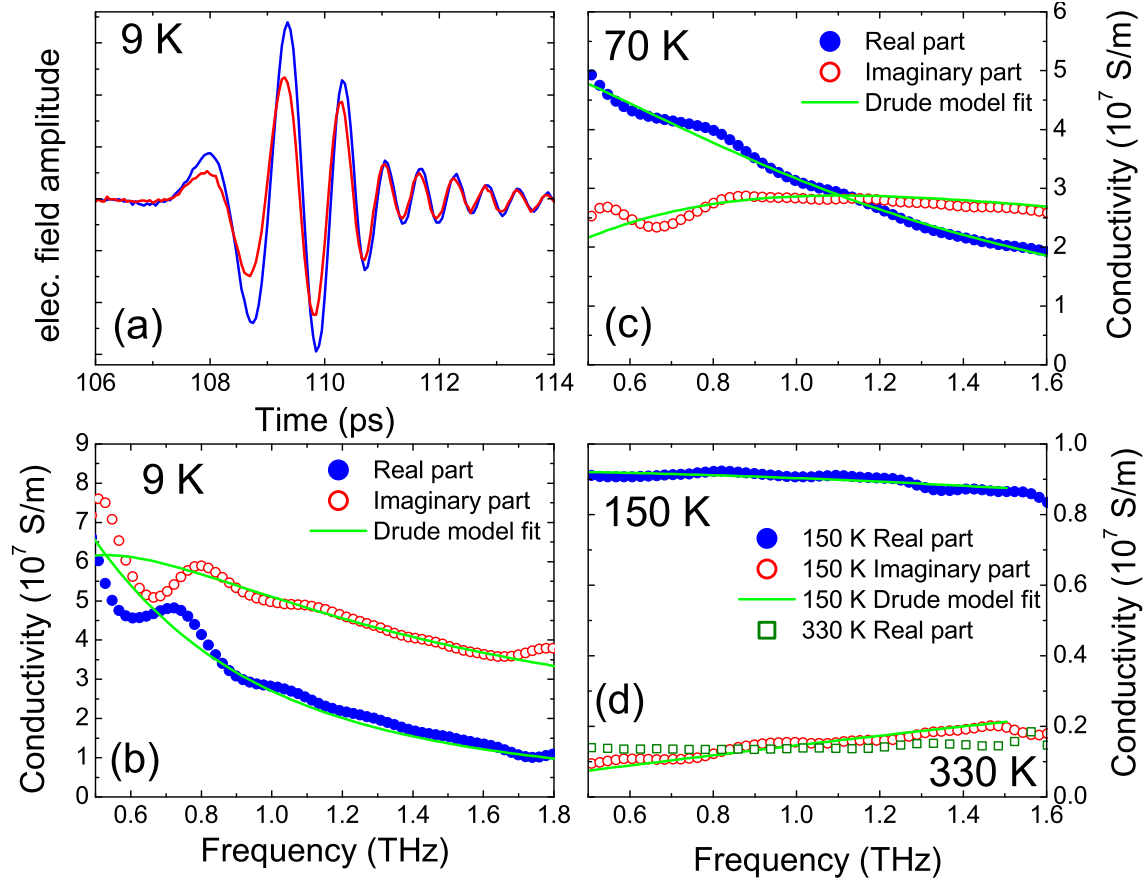}
\caption{\label{fig:thztds}(Color online) (a) Red line: THz pulse transmitted by the CrO$_2$ film in equilibrium state (no pump excitation) at 9 K.  Blue line: the transmitted THz pulse 200 ps after the optical pump at 9 K. (b),(c) Real and imaginary parts of the THz conductivity at 9 and 70 K in equilibrium state.  Symbols: measured conductivity. Solid lines: the Drude model fit.  (d) THz conductivity at 150 K and 330 K.  Only the real part is shown for the 330 K data. Solid lines show the Drude model fit for the 150 K data.}
\end{center}
\end{figure}

\begin{figure}[ht]
\begin{center}
\includegraphics[width=3.7in]{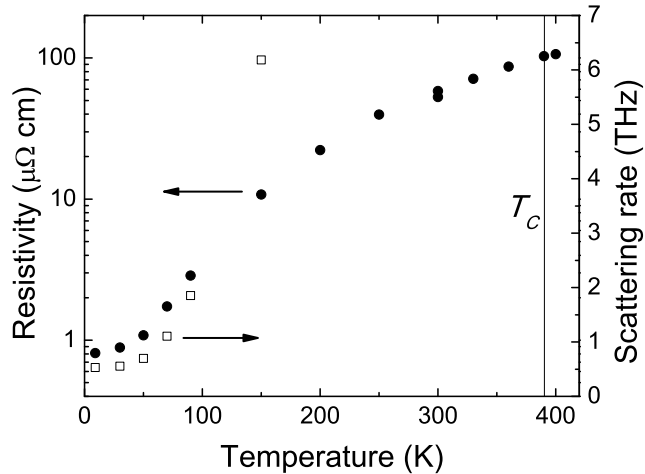}
\caption{\label{fig:drude}(Color online) The temperature dependence of the $dc$ resistivity and the scattering rate $\gamma$ determined by THz TDS.  Drude fit error bars are smaller than the size of the symbols.}
\end{center}
\end{figure}

THz TDS measures the electric field of a THz pulse in time domain.  Figure~\ref{fig:thztds}(a) shows the THz pulse after passing through the CrO$_2$ film at 9 K.  To extract the THz conductivity of the film, a bare TiO$_2$ substrate was used as reference\cite{silwal:092116}.  The sample and reference measurements are Fourier-transformed to obtain the frequency domain spectra $S_{sam}(\omega)$ and $S_{ref}(\omega)$ and compute the amplitude transmission coefficient $\tilde{t}(\omega)=S_{sam}(\omega)/S_{ref}(\omega)$.  We compute the THz optical conductivity $\sigma(\omega)$ from\cite{averitt:1357,silwal:092116}
\begin{equation}
\tilde{t}(\omega)=\frac{\tilde{n}_3+1}{\tilde{n}_3+Z_0\sigma(\omega)d}\exp\left( i\frac{\omega(d_s-d_r)(\tilde{n}_3-1)}{c}\right) ,
\label{eq1}
\end{equation}
where $\tilde{n}_3$ is the THz refractive index of TiO$_2$, $d$ is  the film thickness, $Z_0=377$ $\Omega$ is the free space impedance, and $(d_s-d_r)$ is the difference in thickness between the film and the bare reference substrates. The frequency dependence of the conductivity is well described by the Drude model $\sigma(\omega)=\sigma_0/(1+i\omega/\gamma)$, where $\sigma_0$ is the $dc$ conductivity and $\gamma$ is the electron scattering rate, the parameters whose temperature dependence is determined by least-square fits (Fig.~\ref{fig:thztds}(b-d)).  The real conductivity becomes frequency-independent above 150 K in our THz frequency window and the scattering rate $\gamma$ is not reliably measured.  The temperature dependences of Fig.~\ref{fig:drude} agree well with the published transport and optical conductivity\cite{singley:4126} studies: $\sigma_0$ undergoes a two-order-of-magnitude change, while $\gamma$ also drops precipitously to $\sim0.5$ THz at 9 K.  A similar "collapse of the scattering rate" was found by Singley $et$ $al.$ and is responsible for the low residual resistivity\cite{singley:4126}.

The THz pulse lasts only several picoseconds (Fig.~\ref{fig:thztds}(a)), which allows the measurement of conductivity changes with picosecond time resolution. Figure~\ref{fig:thztds}(a) shows the transmitted THz pulses before and 200 ps after the film is excited by a sub-100-fs optical pump pulse.  The THz probe pulse transmitted before the pump measures the equilibrium conductivity; the probe that passes after the pump measures the conductivity in a non-equilibrium state.  At 9 K, the non-equilibrium THz pulse displays a higher electric field amplitude and a different phase relative to the equilibrium pulse (Fig.~\ref{fig:thztds}(a)).  Above 90 K, the phase difference between equilibrium and non-equilibrium THz probes becomes immeasurable, while the THz amplitude remains higher for the non-equilibrium probe.  The higher amplitude and the different phase of the transmitted THz probe result from a lower conductivity and a higher scattering rate in the non-equilibrium state.  These pump-induced conductivity changes indicate an elevated instantaneous temperature in the evolving non-equilibrium state.  The pump-induced changes are consistent with the findings of Fig.~\ref{fig:drude} that show lower conductivity at higher temperature.  Thus, the effect of the optical pump is a very fast, picosecond-scale heating of the CrO$_2$ film. 

\begin{figure}[ht]
\begin{center}
\includegraphics[width=3in]{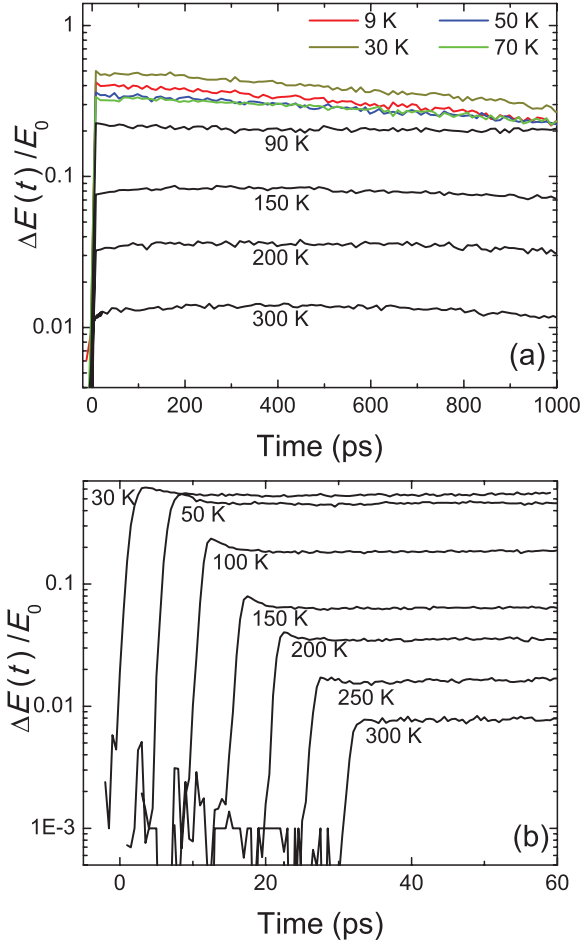}
\caption{\label{fig:trts}(Color online) (a) TRTS spectra - the time-resolved measurement of the ultrafast change in the peak transmitted THz electric field at various temperatures.  (b) TRTS spectra similar to (a), but zoomed in on the first 30-60 ps of the photoinduced response and recorded with a shorter time step.  The spectra were shifted horizontally for clarity.}
\end{center}
\end{figure}

By varying the arrival time of the THz probe relative to the optical pump, we record the evolution of pump-induced conductivity changes.  For simplicity, we only measure the change in the peak transmitted THz field instead of recording the full THz pulse.  Figure~\ref{fig:trts} shows the measured relative change $\Delta E(t)/E_0$ as a function of time delay between the pump and probe; $E_0$ is the peak THz field in the absence of the pump excitation.

Since conductivity $\sigma(\omega)$ is (almost) independent of frequency $\omega$ above 150 K, we can take $\sigma(\omega)\approx\sigma_0$ and relate the change $\Delta E(t)/E_0$ to the frequency-independent pump-induced change  $\Delta\sigma$ as 
\begin{equation}
\frac{\Delta E(t)}{E_0}=\frac{-Z_0d\Delta\sigma}{1+\tilde{n}_3+Z_0d\sigma_0}.
\label{eq2}
\end{equation}
On the right hand side of Eq. (\ref{eq2}), only $\Delta\sigma$ contains the effect of the optical pump.  All other quantities characterize the equilibrium state.  Thus, the time-evolution of $\Delta E(t)/E_0$ reflects the time-evolution of $\Delta\sigma$.  A higher transmitted THz field (positive $\Delta E$) indicates a drop in conductivity (negative $\Delta\sigma$).

Figure~\ref{fig:trts}(a) shows two vastly different time scales in the conductivity response to the optical pump.  A fast step-like rise in $\Delta E(t)$ is followed by a much slower evolution, as $\Delta E(t)$ reaches a broad maximum (near 400 ps at 300 K in Fig.~\ref{fig:trts}(a)) and then begins a slow recovery of its equilibrium value.  Below 250 K, the initial rise in $\Delta E(t)$ is followed by a fast shallow drop (Fig.~\ref{fig:trts}(b)) before the broad maximum and the recovery of equilibrium.  The broad maximum in $\Delta E(t)$ is found at all temperatures down to 70 K but becomes a lot less pronounced below 150 K (Fig.~\ref{fig:trts}(a)). 

\section{Discussion of the results}
The dynamics of $\Delta E(t)$ reflects the relaxation of the pump excitation energy.  The 1.55-eV pump photons are absorbed by the transitions in the majority channel, as the gap in the minority channel exceeds 2 eV.  The optical pump creates a highly excited non-thermal electron population.  The evolution of this excited state is usually described in terms of a fast ($\sim 100$ fs) electron thermalization at an elevated temperature, which is followed by the electron-phonon relaxation and the equilibration of electronic and phonon temperatures\cite{talbayev:340,averitt:1357,groeneveld:11433,allen:1460,costa:104407}.  Zhang $et$ $al.$ found that at 300 K in CrO$_2$, the electron and phonon temperatures reach an equilibrium in about 2-3 ps\cite{zhang:064414}, which is consistent with our data.  The fast shallow drop in $\Delta E(t)$ at $T\leq250$ K corresponds to a slight recovery of conductivity as energy is transferred from electrons to phonons (Fig.~\ref{fig:trts}(b)).  The absence of this conductivity recovery feature at high temperature indicates that the phonon scattering and the phonon temperature gain relative importance in the conductivity dynamics.  With or without the slight recovery, we interpret the initial ($\leq 5$ ps) dynamics in $\Delta E(t)$ as the electron-phonon relaxation, after which elevated and equal electron and phonon temperatures are established, leading to higher electron-electron and electron-phonon scattering rates.

\begin{figure}[ht]
\begin{center}
\includegraphics[width=3in]{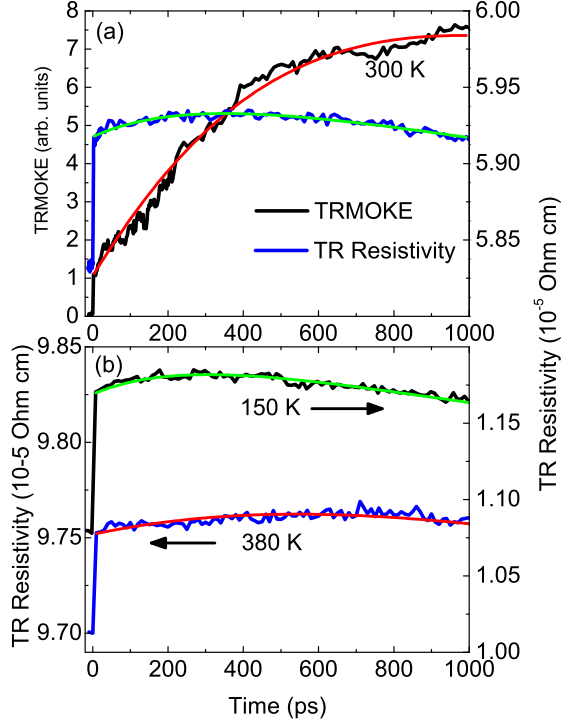}
\caption{\label{fig:trmoke}(Color online) (a) TRTS and TRMOKE spectra at room temperature. Solid lines are fits to the two-temperature model. (b) TRTS spectra at 150 K and 380 K.  The solid lines are fits to the two-temperature model.}
\end{center}
\end{figure}

Why does the conductivity $\Delta\sigma(t)$ continue to drop (the resistivity $\Delta\rho(t)$ continue to rise) after the initial dynamics?  Another process that contributes to resistivity is the spin-flip scattering.  The evidence for the spin temperature evolution in CrO$_2$ after the optical pump is provided by the magneto-optical Kerr effect\cite{zhang:064414,mueller:56} (MOKE), which refers to a change in the polarization state of reflected light and is proportional to the material's magnetization.  In time-resolved MOKE (TRMOKE), the pump-induced change in magnetization is recorded\cite{zhang:177402,carpene:174437,munzenberg:020412,zhao:207205,talbayev:014417,talbayev:182501}.  In CrO$_2$, a slow demagnetization over hundreds of picoseconds follows the optical pump excitation, as the spin temperature rises due to the spin-lattice coupling\cite{zhang:064414,mueller:56}.  Figure~\ref{fig:trmoke}(a) shows a room-temperature TRMOKE measurement in which we recorded the pump-induced polarization rotation of an optical 1.55-eV probe  pulse; we observe a fast initial jump and a much slower increase over 1 ns.  The comparison with published data\cite{zhang:064414,mueller:56} shows that the slow TRMOKE dynamics reflects the demagnetization as the spin temperature equilibrates with the electron and phonon temperature via the spin-lattice coupling.  The room-temperature spin-lattice relaxation time was measured by Zhang $et$ $al.$\cite{zhang:064414} to be $\sim 400$ ps.  Thus, the broad maximum in $\Delta E(t)$ results from the rise in the spin temperature and the corresponding increase in the spin-flip scattering. 

To quantify the effect of the spin fluctuations on resistivity, we turn to the two-temperature model of the coupled electron-lattice and spin dynamics\cite{zhang:064414,mueller:56}.  This model is applicable for the long time-scale ($t>5$ ps) dynamics, when the electron and phonon temperatures can be taken as equal and described by a single electron-phonon temperature $T_{ep}$.  The spin system is described by the spin temperature $T_s$.  The spin and electron-phonon dynamics are described by a pair of differential equations 
\begin{eqnarray}
c_{ep}\frac{\partial T_{ep}}{\partial t}=-g\left( T_{ep}-T_s\right) - \beta\left( \left[ 1-exp(-t/\tau)\right] T_{ep} - T_b\right) \label{eq:ttm1},\\
c_s\frac{\partial T_s}{\partial t}=-g\left( T_s-T_{ep}\right),
\label{eq:ttm2}
\end{eqnarray}
where $c_{ep}$ and $c_s$ are electron-phonon and spin specific heats and $g$ is the spin-lattice coupling constant\cite{zhang:064414}.  The specific heats $c_{ep}$ and $c_s$ are taken as temperature indepedent under the assumption of a small pump-induced temperature change in the electron-phonon and spin systems.  The last term in (\ref{eq:ttm1}) describes the cooling of the electron-phonon system by the diffusion of energy into the substrate, whose temperature $T_b$ is taken as constant and equal to the equilibrium temperature of the measurement.  The cooling is proportional to the temperature difference $(T_{ep}-T_b)$ and is parametrized by a constant $\beta$.  The exponential that multiplies $T_{ep}$ in the last term of Eq. (\ref{eq:ttm1}) accounts for the gradual "turning on" of the cooling as the energy deposited within the optical absorption depth of the pump wavelength diffuses through the film thickness to the substrate side of the film.  We estimate the optical penetration depth to be 17 nm using the optical constants measured by Stewart $et$ $al.$\cite{stewart:144414}

\begin{figure}[ht]
\begin{center}
\includegraphics[width=3in]{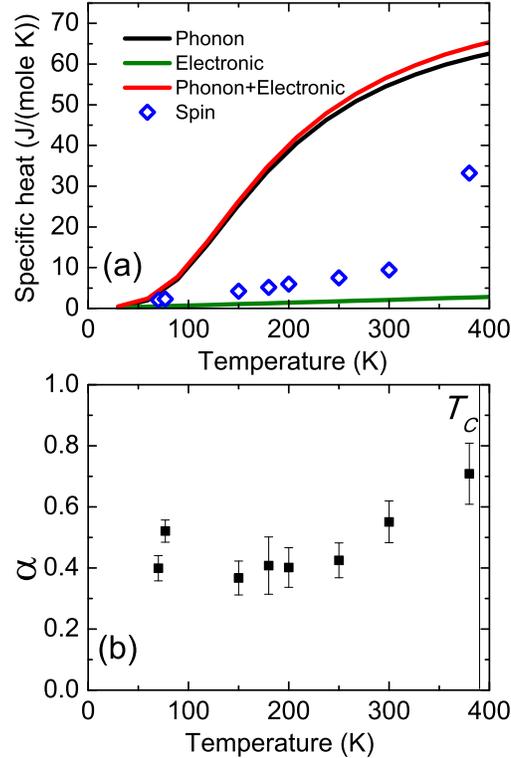}
\caption{\label{fig:alpha}(Color online) (a) The computed electron-phonon and spin specific heat.  (b) Temperature dependence of the parameter $\alpha$, which quantifies the relative contribution of the spin temperature change to the photoinduced resistivity change.  The vertical line indicates the Curie temperature $T_c=390$ K.}
\end{center}
\end{figure}

The electron-phonon specific heat $c_{ep}$ consists of the electronic and phonon specific heats.  We compute the electronic contribution as\cite{lewis:10253} $c_e=\gamma T$ with $\gamma=7$ mJ K$^{-2}$mole$^{-1}$.  We compute the acoustic phonon specific heat in the Debye model with $\theta_D=593$ K\cite{coey:8345}.  The optical phonon specific heat is computed in the Einstein model using the frequency $\omega_{opt}=450$ cm$^{-1}$ to represent the branches of the optical phonon spectrum\cite{iliev:33}.  For the spin specific heat, we use the mean-field value\cite{zhang:064414} $c_s(T)=-\alpha\frac{\partial M^2}{\partial T}$, where $\alpha=3SRT_C/[2(S+1)M(0)^2]$, $S=2$ for CrO$_2$, and $M(0)$ is the saturation magnetization at low temperature.  We use the magnetometry measurements of Li $et$ $al.$\cite{li:5585} to determine $M(0)^2$ and $\frac{\partial M^2}{\partial T}$.  The computed electron-phonon and spin specific heats are shown in Fig.~\ref{fig:alpha}(a).  

\begin{figure}[ht]
\begin{center}
\includegraphics[width=3in]{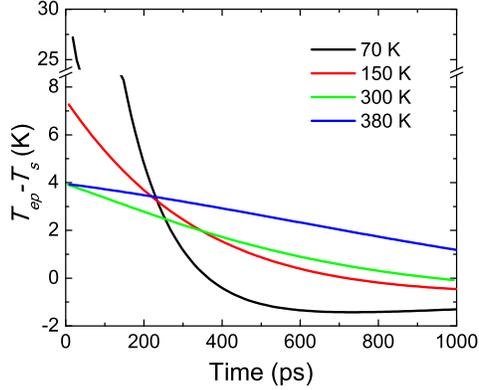}
\caption{\label{fig:tempdiff}(Color online) Time evolution of the difference in the instantaneous electron-lattice and spin temperature computed in the two-temperature model.  Colors correspond to different equilibrium temperatures.}
\end{center}
\end{figure}

Next, we use the two-temperature model (Eqs. (\ref{eq:ttm1},\ref{eq:ttm2})) to fit the experimental TRTS data.  We compute the time-resolved ultrafast change in resistivity as
\begin{equation}
\rho(t)=\rho_0(T)+\frac{\partial\rho_0(T)}{\partial T}\left[ (1-\alpha)\Delta T_{ep}(t)+\alpha\Delta T_s(t)\right],
\label{eq:alpha}
\end{equation}
where $\rho_0(T)$ is the equilibrium temperature-dependent resistivity, $\partial\rho_0(T)/\partial T$ is the slope of $\rho_0(T)$, and both are taken from Fig.~\ref{fig:drude}.  The parameter $\alpha$ describes the relative importance of the evolving photoinduced change in electron-phonon and spin temperatures, $\Delta T_{ep}(t)$ and $\Delta\T_s(t)$, in the determination of the photoinduced resistivity $\rho(t)$.  We set the initial spin temperature as equal to the equilibrium temperature of the measurement, $T_s(0)=T_b$.  At 300 K, we have both TRMOKE and TRTS data and we fit both of them simultaneously by assuming that the TRMOKE angle is proportional to $\Delta T_s(t)$.  Fitting both data sets with the same model parameters allows us to determine the parameters $g$ and $\alpha$ simultaneously.  We find the value of the spin-lattice coupling constant $g=0.011$ J/(mole K ps), which compares well to $g=0.018$ J/(mole K ps) deduced from the data of Zhang $et$ $al.$\cite{zhang:064414}  The fractional photoinduced magnetization change was estimated to be $0.6\%$ at 1000 ps time delay (Fig.~\ref{fig:trmoke}(a)).  To fit the TRTS data at the other temperatures, we consider the spin lattice coupling $g$ as independent of temperature\cite{averitt:017401}.  Figure~\ref{fig:alpha}(b) shows the obtained temperature dependence of the fitting parameter $\alpha$.  We tested how stable our fitting procedure was under the variation of the fitting parameters $\beta$ and $\tau$ that describe the cooling of the electron-phonon system by energy diffusion into the substrate via the empirical last term in Eq. (\ref{eq:ttm1}).  We varied $\tau$ between 10 ps and 10000 ps and obtained similar quality fits to the experimental data. While the fitting parameter $\beta$ needed to be adjusted by a significant amount to accommodate the large range of $\tau$, the fitted values of $\alpha$ changed only very slightly, as reflected by its standard deviation (the error bars) reported in Fig.~\ref{eq:alpha}(b). Figure~\ref{fig:tempdiff} shows the computed evolution of the temperature difference ($T_{ep}(t)-T_s(t)$) and indicates that the photoinduced instantaneous change in electron and spin temperatures is small at all but the lowest ($T\leq77$ K) temperatures in our measurement.

Boltzmann transport theory describes the electric current by summing up the electron velocities over the occupied quasiparticle states.  The resistivity is caused by quasiparticle scattering between crystal momentum states, and is proportional to the electron scattering rate $\gamma$: $\rho_0(T)=\gamma/(\epsilon_0 \omega_p^2)$, with $\epsilon_0$ being the permittivity of free space and $\omega_p$ being the plasma frequency.  When the various scattering mechanisms are independent, they are combined using Matthiessen's rule as $\gamma(T)=\gamma_{ep}(T)+\gamma_{s}(T)$, where $\gamma_{ep}(T)$ includes the electron-phonon and electron-electron scattering and $\gamma_s(T)$ describes the spin-flip scattering.  Since the plasma frequency $\omega_p$ in CrO$_2$ is independent of temperature\cite{stewart:144414} below $T_C$, the photoinduced change in resistivity happens because the elevated instantaneous electron-phonon and spin temperatures modify the respective scattering rates, $\gamma_{ep}(T_{ep})$ and $\gamma_{s}(T_s)$.  The parameter $\alpha$ introduced in Eq. (\ref{eq:alpha}) quantifies the relative contribution of the spin-flip scattering to all scattering events.  According to Fig.~\ref{fig:alpha}(b), the spin-flip processes do not dominate the electron scattering in the wide temperature range 70-300 K where the most of the temperature-induced change in resistivity occurs below $T_C$ (Fig.~\ref{fig:drude}).  The exception are the temperatures above and near $T_C$, where $\alpha$ reaches about 70\% at 380 K. 

It is instructive to compare our findings with a study by Averitt et al.\citep{averitt:017401} of the double exchange manganites La$_{0.7}$Ca(Sr)$_{0.3}$MnO$_3$.  The manganites exhibit an ultrafast conductivity response with two distinct time scales, with the short $\sim2$ ps scale attributed to electron-phonon relaxation, and the longer scale of tens of ps attributed to spin-lattice relaxation\citep{averitt:017401}.  The phonon-induced conductivity change $\Delta\rho(\Delta T_{ep})$ dominates at low temperature ($T<0.5 T_C$), while the spin-fluctuation-induced $\Delta\rho(\Delta T_s)$ dominates close to $T_C$.  This behavior is similar to our findings in CrO$_2$.  Spin fluctuations are greatly enhanced near $T_C$, and the relative importance of the spin-flip scattering rises in both materials.  However, there is also a significant difference between CrO$_2$ and the manganites: in the manganites, the resistivity is highly sensitive to temperature near $T_C$ and even exhibits a metal-insulator transition driven by the double-exchange physics\cite{schiffer:3336}.  In CrO$_2$, the resistivity is featureless near the ferromagnetic phase transition\cite{gupta:6073,li:5585} (Fig.~\ref{fig:drude}), despite the relative enhancement of the spin-flip scattering close to $T_C$.  Our results point to a significant disconnect between charge transport and magnetic order, which conflicts with the double exchange scenario of magnetism in CrO$_2$.  Other evidence for such conflict is provided by the spectroscopic ellipsometry study of CrO$_2$ films by Stewart $et$ $al.$\cite{stewart:144414}, who found that the Drude plasma frequency and the effective number of carriers remain constant across the ferromagnetic phase transition.  By contrast, in the double-exchange manganites, a significant transfer of the spectral weight is found from high to low energy and the Drude response grows in strength as the temperature is lowered below $T_C$\cite{okimoto:109,kim:1517,lee:5251}.

\section{Summary}
We have presented a TRTS study of the half-metallic ferromagnet CrO$_2$, where the ultrafast resistivity response is governed by the electron-phonon and spin-lattice relaxation.  In the first 2-5 ps after the optical pump pulse, we observe a step-like change in the photoinduced time-resolved resistivity $\rho(t)$, which we ascribe to the establishment of an elevated electron and phonon temperature (Fig.~\ref{fig:trmoke}).  The fast step-like feature is followed by a continued slow rise in $\rho(t)$ before the recovery of the equilibrium state begins.  The slow rise in $\rho(t)$ after the initial fast dynamics can be explained as the heating of the spin system via spin-lattice coupling.  We use a two-temperature model of coupled electron-phonon and spin dynamics to quantify the contribution of the spin temperature change to the change in resistivity. As shown in Fig.~\ref{fig:alpha}(b), the spin fluctuations provide the dominant contribution to photoinduced resistivity only near $T_C$.  Below room temperature, the electron-phonon and electron-electron scattering dominates.  This finding should provide further guidance to theoretical descriptions of electronic transport in CrO$_2$. Many of the past theoretical models neglected either the spin-fluctuation or the electron and phonon scattering, while our results show that both must taken into account.

The work at Tulane was supported by the Louisiana Board of Regents through the Board of Regents Support Fund contract number LEQSF(2012-15)-RD-A-23 and through the LA EPSCoR contract number LEQSF-EPS(2014)-PFUND-378.  The work at the University of Alabama was supported by NSF Grant No. DMR-0706280.


\begin{thebibliography}{41}
\expandafter\ifx\csname natexlab\endcsname\relax\def\natexlab#1{#1}\fi
\expandafter\ifx\csname bibnamefont\endcsname\relax
  \def\bibnamefont#1{#1}\fi
\expandafter\ifx\csname bibfnamefont\endcsname\relax
  \def\bibfnamefont#1{#1}\fi
\expandafter\ifx\csname citenamefont\endcsname\relax
  \def\citenamefont#1{#1}\fi
\expandafter\ifx\csname url\endcsname\relax
  \def\url#1{\texttt{#1}}\fi
\expandafter\ifx\csname urlprefix\endcsname\relax\def\urlprefix{URL }\fi
\providecommand{\bibinfo}[2]{#2}
\providecommand{\eprint}[2][]{\url{#2}}

\bibitem[{\citenamefont{Schwarz}(1986)}]{schwarz:l211}
\bibinfo{author}{\bibfnamefont{K.}~\bibnamefont{Schwarz}},
  \bibinfo{journal}{Journal of Physics F: Metal Physics}
  \textbf{\bibinfo{volume}{16}}, \bibinfo{pages}{L211} (\bibinfo{year}{1986}),
  \urlprefix\url{http://stacks.iop.org/0305-4608/16/i=9/a=002}.

\bibitem[{\citenamefont{K\"amper et~al.}(1987)\citenamefont{K\"amper, Schmitt,
  G\"untherodt, Gambino, and Ruf}}]{kamper:2788}
\bibinfo{author}{\bibfnamefont{K.~P.} \bibnamefont{K\"amper}},
  \bibinfo{author}{\bibfnamefont{W.}~\bibnamefont{Schmitt}},
  \bibinfo{author}{\bibfnamefont{G.}~\bibnamefont{G\"untherodt}},
  \bibinfo{author}{\bibfnamefont{R.~J.} \bibnamefont{Gambino}},
  \bibnamefont{and} \bibinfo{author}{\bibfnamefont{R.}~\bibnamefont{Ruf}},
  \bibinfo{journal}{Phys. Rev. Lett.} \textbf{\bibinfo{volume}{59}},
  \bibinfo{pages}{2788} (\bibinfo{year}{1987}),
  \urlprefix\url{http://link.aps.org/doi/10.1103/PhysRevLett.59.2788}.

\bibitem[{\citenamefont{Wiesendanger et~al.}(1990)\citenamefont{Wiesendanger,
  G\"untherodt, G\"untherodt, Gambino, and Ruf}}]{guntherodt:247}
\bibinfo{author}{\bibfnamefont{R.}~\bibnamefont{Wiesendanger}},
  \bibinfo{author}{\bibfnamefont{H.-J.} \bibnamefont{G\"untherodt}},
  \bibinfo{author}{\bibfnamefont{G.}~\bibnamefont{G\"untherodt}},
  \bibinfo{author}{\bibfnamefont{R.~J.} \bibnamefont{Gambino}},
  \bibnamefont{and} \bibinfo{author}{\bibfnamefont{R.}~\bibnamefont{Ruf}},
  \bibinfo{journal}{Phys. Rev. Lett.} \textbf{\bibinfo{volume}{65}},
  \bibinfo{pages}{247} (\bibinfo{year}{1990}),
  \urlprefix\url{http://link.aps.org/doi/10.1103/PhysRevLett.65.247}.

\bibitem[{\citenamefont{Soulen et~al.}(1998)\citenamefont{Soulen, Byers,
  Osofsky, Nadgorny, Ambrose, Cheng, Broussard, Tanaka, Nowak, Moodera
  et~al.}}]{soulen:85}
\bibinfo{author}{\bibfnamefont{R.~J.} \bibnamefont{Soulen}},
  \bibinfo{author}{\bibfnamefont{J.~M.} \bibnamefont{Byers}},
  \bibinfo{author}{\bibfnamefont{M.~S.} \bibnamefont{Osofsky}},
  \bibinfo{author}{\bibfnamefont{B.}~\bibnamefont{Nadgorny}},
  \bibinfo{author}{\bibfnamefont{T.}~\bibnamefont{Ambrose}},
  \bibinfo{author}{\bibfnamefont{S.~F.} \bibnamefont{Cheng}},
  \bibinfo{author}{\bibfnamefont{P.~R.} \bibnamefont{Broussard}},
  \bibinfo{author}{\bibfnamefont{C.~T.} \bibnamefont{Tanaka}},
  \bibinfo{author}{\bibfnamefont{J.}~\bibnamefont{Nowak}},
  \bibinfo{author}{\bibfnamefont{J.~S.} \bibnamefont{Moodera}},
  \bibnamefont{et~al.}, \bibinfo{journal}{Science}
  \textbf{\bibinfo{volume}{282}}, \bibinfo{pages}{85} (\bibinfo{year}{1998}),
  \eprint{http://www.sciencemag.org/content/282/5386/85.full.pdf},
  \urlprefix\url{http://www.sciencemag.org/content/282/5386/85.abstract}.

\bibitem[{\citenamefont{Anguelouch et~al.}(2001)\citenamefont{Anguelouch,
  Gupta, Xiao, Abraham, Ji, Ingvarsson, and Chien}}]{anguelouch:180408}
\bibinfo{author}{\bibfnamefont{A.}~\bibnamefont{Anguelouch}},
  \bibinfo{author}{\bibfnamefont{A.}~\bibnamefont{Gupta}},
  \bibinfo{author}{\bibfnamefont{G.}~\bibnamefont{Xiao}},
  \bibinfo{author}{\bibfnamefont{D.~W.} \bibnamefont{Abraham}},
  \bibinfo{author}{\bibfnamefont{Y.}~\bibnamefont{Ji}},
  \bibinfo{author}{\bibfnamefont{S.}~\bibnamefont{Ingvarsson}},
  \bibnamefont{and} \bibinfo{author}{\bibfnamefont{C.~L.} \bibnamefont{Chien}},
  \bibinfo{journal}{Phys. Rev. B} \textbf{\bibinfo{volume}{64}},
  \bibinfo{pages}{180408} (\bibinfo{year}{2001}),
  \urlprefix\url{http://link.aps.org/doi/10.1103/PhysRevB.64.180408}.

\bibitem[{\citenamefont{Parker et~al.}(2002)\citenamefont{Parker, Watts,
  Ivanov, and Xiong}}]{parker:196601}
\bibinfo{author}{\bibfnamefont{J.~S.} \bibnamefont{Parker}},
  \bibinfo{author}{\bibfnamefont{S.~M.} \bibnamefont{Watts}},
  \bibinfo{author}{\bibfnamefont{P.~G.} \bibnamefont{Ivanov}},
  \bibnamefont{and} \bibinfo{author}{\bibfnamefont{P.}~\bibnamefont{Xiong}},
  \bibinfo{journal}{Phys. Rev. Lett.} \textbf{\bibinfo{volume}{88}},
  \bibinfo{pages}{196601} (\bibinfo{year}{2002}),
  \urlprefix\url{http://link.aps.org/doi/10.1103/PhysRevLett.88.196601}.

\bibitem[{\citenamefont{Chamberland}(1977)}]{chamberland:1}
\bibinfo{author}{\bibfnamefont{B.~L.} \bibnamefont{Chamberland}},
  \bibinfo{journal}{Critical Reviews in Solid State and Materials Sciences}
  \textbf{\bibinfo{volume}{7}}, \bibinfo{pages}{1} (\bibinfo{year}{1977}),
  \eprint{http://dx.doi.org/10.1080/10408437708243431},
  \urlprefix\url{http://dx.doi.org/10.1080/10408437708243431}.

\bibitem[{\citenamefont{Korotin et~al.}(1998)\citenamefont{Korotin, Anisimov,
  Khomskii, and Sawatzky}}]{korotin:4305}
\bibinfo{author}{\bibfnamefont{M.~A.} \bibnamefont{Korotin}},
  \bibinfo{author}{\bibfnamefont{V.~I.} \bibnamefont{Anisimov}},
  \bibinfo{author}{\bibfnamefont{D.~I.} \bibnamefont{Khomskii}},
  \bibnamefont{and} \bibinfo{author}{\bibfnamefont{G.~A.}
  \bibnamefont{Sawatzky}}, \bibinfo{journal}{Phys. Rev. Lett.}
  \textbf{\bibinfo{volume}{80}}, \bibinfo{pages}{4305} (\bibinfo{year}{1998}),
  \urlprefix\url{http://link.aps.org/doi/10.1103/PhysRevLett.80.4305}.

\bibitem[{\citenamefont{Anwar and Aarts}(2013)}]{anwar:085123}
\bibinfo{author}{\bibfnamefont{M.~S.} \bibnamefont{Anwar}} \bibnamefont{and}
  \bibinfo{author}{\bibfnamefont{J.}~\bibnamefont{Aarts}},
  \bibinfo{journal}{Phys. Rev. B} \textbf{\bibinfo{volume}{88}},
  \bibinfo{pages}{085123} (\bibinfo{year}{2013}),
  \urlprefix\url{http://link.aps.org/doi/10.1103/PhysRevB.88.085123}.

\bibitem[{\citenamefont{Lewis et~al.}(1997)\citenamefont{Lewis, Allen, and
  Sasaki}}]{lewis:10253}
\bibinfo{author}{\bibfnamefont{S.~P.} \bibnamefont{Lewis}},
  \bibinfo{author}{\bibfnamefont{P.~B.} \bibnamefont{Allen}}, \bibnamefont{and}
  \bibinfo{author}{\bibfnamefont{T.}~\bibnamefont{Sasaki}},
  \bibinfo{journal}{Phys. Rev. B} \textbf{\bibinfo{volume}{55}},
  \bibinfo{pages}{10253} (\bibinfo{year}{1997}),
  \urlprefix\url{http://link.aps.org/doi/10.1103/PhysRevB.55.10253}.

\bibitem[{\citenamefont{Coey and Venkatesan}(2002)}]{coey:8345}
\bibinfo{author}{\bibfnamefont{J.~M.~D.} \bibnamefont{Coey}} \bibnamefont{and}
  \bibinfo{author}{\bibfnamefont{M.}~\bibnamefont{Venkatesan}},
  \bibinfo{journal}{Journal of Applied Physics} \textbf{\bibinfo{volume}{91}},
  \bibinfo{pages}{8345} (\bibinfo{year}{2002}),
  \urlprefix\url{http://scitation.aip.org/content/aip/journal/jap/91/10/10.1063/1.1447879}.

\bibitem[{\citenamefont{Barry et~al.}(1998)\citenamefont{Barry, Coey, Ranno,
  and Ounadjela}}]{barry:7166}
\bibinfo{author}{\bibfnamefont{A.}~\bibnamefont{Barry}},
  \bibinfo{author}{\bibfnamefont{J.~M.~D.} \bibnamefont{Coey}},
  \bibinfo{author}{\bibfnamefont{L.}~\bibnamefont{Ranno}}, \bibnamefont{and}
  \bibinfo{author}{\bibfnamefont{K.}~\bibnamefont{Ounadjela}},
  \bibinfo{journal}{Journal of Applied Physics} \textbf{\bibinfo{volume}{83}},
  \bibinfo{pages}{7166} (\bibinfo{year}{1998}),
  \urlprefix\url{http://scitation.aip.org/content/aip/journal/jap/83/11/10.1063/1.367791}.

\bibitem[{\citenamefont{Gupta et~al.}(2000)\citenamefont{Gupta, Li, and
  Xiao}}]{gupta:6073}
\bibinfo{author}{\bibfnamefont{A.}~\bibnamefont{Gupta}},
  \bibinfo{author}{\bibfnamefont{X.~W.} \bibnamefont{Li}}, \bibnamefont{and}
  \bibinfo{author}{\bibfnamefont{G.}~\bibnamefont{Xiao}},
  \bibinfo{journal}{Journal of Applied Physics} \textbf{\bibinfo{volume}{87}},
  \bibinfo{pages}{6073} (\bibinfo{year}{2000}),
  \urlprefix\url{http://scitation.aip.org/content/aip/journal/jap/87/9/10.1063/1.372616}.

\bibitem[{\citenamefont{Watts et~al.}(2000)\citenamefont{Watts, Wirth, von
  Moln\'ar, Barry, and Coey}}]{watts:9621}
\bibinfo{author}{\bibfnamefont{S.~M.} \bibnamefont{Watts}},
  \bibinfo{author}{\bibfnamefont{S.}~\bibnamefont{Wirth}},
  \bibinfo{author}{\bibfnamefont{S.}~\bibnamefont{von Moln\'ar}},
  \bibinfo{author}{\bibfnamefont{A.}~\bibnamefont{Barry}}, \bibnamefont{and}
  \bibinfo{author}{\bibfnamefont{J.~M.~D.} \bibnamefont{Coey}},
  \bibinfo{journal}{Phys. Rev. B} \textbf{\bibinfo{volume}{61}},
  \bibinfo{pages}{9621} (\bibinfo{year}{2000}),
  \urlprefix\url{http://link.aps.org/doi/10.1103/PhysRevB.61.9621}.

\bibitem[{\citenamefont{Yanagihara and Salamon}(2002)}]{yanagihara:187201}
\bibinfo{author}{\bibfnamefont{H.}~\bibnamefont{Yanagihara}} \bibnamefont{and}
  \bibinfo{author}{\bibfnamefont{M.~B.} \bibnamefont{Salamon}},
  \bibinfo{journal}{Phys. Rev. Lett.} \textbf{\bibinfo{volume}{89}},
  \bibinfo{pages}{187201} (\bibinfo{year}{2002}),
  \urlprefix\url{http://link.aps.org/doi/10.1103/PhysRevLett.89.187201}.

\bibitem[{\citenamefont{Branford et~al.}(2009)\citenamefont{Branford, Yates,
  Barkhoudarov, Moore, Morrison, Magnus, Miyoshi, Sousa, Conde, Silvestre
  et~al.}}]{branford:227201}
\bibinfo{author}{\bibfnamefont{W.~R.} \bibnamefont{Branford}},
  \bibinfo{author}{\bibfnamefont{K.~A.} \bibnamefont{Yates}},
  \bibinfo{author}{\bibfnamefont{E.}~\bibnamefont{Barkhoudarov}},
  \bibinfo{author}{\bibfnamefont{J.~D.} \bibnamefont{Moore}},
  \bibinfo{author}{\bibfnamefont{K.}~\bibnamefont{Morrison}},
  \bibinfo{author}{\bibfnamefont{F.}~\bibnamefont{Magnus}},
  \bibinfo{author}{\bibfnamefont{Y.}~\bibnamefont{Miyoshi}},
  \bibinfo{author}{\bibfnamefont{P.~M.} \bibnamefont{Sousa}},
  \bibinfo{author}{\bibfnamefont{O.}~\bibnamefont{Conde}},
  \bibinfo{author}{\bibfnamefont{A.~J.} \bibnamefont{Silvestre}},
  \bibnamefont{et~al.}, \bibinfo{journal}{Phys. Rev. Lett.}
  \textbf{\bibinfo{volume}{102}}, \bibinfo{pages}{227201}
  (\bibinfo{year}{2009}),
  \urlprefix\url{http://link.aps.org/doi/10.1103/PhysRevLett.102.227201}.

\bibitem[{\citenamefont{Suzuki and Tedrow}(1998)}]{suzuki:11597}
\bibinfo{author}{\bibfnamefont{K.}~\bibnamefont{Suzuki}} \bibnamefont{and}
  \bibinfo{author}{\bibfnamefont{P.~M.} \bibnamefont{Tedrow}},
  \bibinfo{journal}{Phys. Rev. B} \textbf{\bibinfo{volume}{58}},
  \bibinfo{pages}{11597} (\bibinfo{year}{1998}),
  \urlprefix\url{http://link.aps.org/doi/10.1103/PhysRevB.58.11597}.

\bibitem[{\citenamefont{Talbayev et~al.}(2012)\citenamefont{Talbayev, Chia,
  Trugman, Zhu, and Taylor}}]{talbayev:340}
\bibinfo{author}{\bibfnamefont{D.}~\bibnamefont{Talbayev}},
  \bibinfo{author}{\bibfnamefont{E.~E.~M.} \bibnamefont{Chia}},
  \bibinfo{author}{\bibfnamefont{S.~A.} \bibnamefont{Trugman}},
  \bibinfo{author}{\bibfnamefont{J.-X.} \bibnamefont{Zhu}}, \bibnamefont{and}
  \bibinfo{author}{\bibfnamefont{A.~J.} \bibnamefont{Taylor}},
  \bibinfo{journal}{IEEE J. Sel. Topics. Quantum Elec.}
  \textbf{\bibinfo{volume}{18}}, \bibinfo{pages}{340 } (\bibinfo{year}{2012}).

\bibitem[{\citenamefont{Burch et~al.}(2008)\citenamefont{Burch, Chia, Talbayev,
  Sales, Mandrus, Taylor, and Averitt}}]{burch:026409}
\bibinfo{author}{\bibfnamefont{K.~S.} \bibnamefont{Burch}},
  \bibinfo{author}{\bibfnamefont{E.~E.~M.} \bibnamefont{Chia}},
  \bibinfo{author}{\bibfnamefont{D.}~\bibnamefont{Talbayev}},
  \bibinfo{author}{\bibfnamefont{B.~C.} \bibnamefont{Sales}},
  \bibinfo{author}{\bibfnamefont{D.}~\bibnamefont{Mandrus}},
  \bibinfo{author}{\bibfnamefont{A.~J.} \bibnamefont{Taylor}},
  \bibnamefont{and} \bibinfo{author}{\bibfnamefont{R.~D.}
  \bibnamefont{Averitt}}, \bibinfo{journal}{Phys. Rev. Lett.}
  \textbf{\bibinfo{volume}{100}}, \bibinfo{pages}{026409}
  (\bibinfo{year}{2008}),
  \urlprefix\url{http://link.aps.org/doi/10.1103/PhysRevLett.100.026409}.

\bibitem[{\citenamefont{Costa et~al.}(2015)\citenamefont{Costa, Huisman,
  Mikhaylovskiy, Razdolski, Ventura, Teixeira, Schmool, Kakazei, Cardoso,
  Freitas et~al.}}]{costa:104407}
\bibinfo{author}{\bibfnamefont{J.~D.} \bibnamefont{Costa}},
  \bibinfo{author}{\bibfnamefont{T.~J.} \bibnamefont{Huisman}},
  \bibinfo{author}{\bibfnamefont{R.~V.} \bibnamefont{Mikhaylovskiy}},
  \bibinfo{author}{\bibfnamefont{I.}~\bibnamefont{Razdolski}},
  \bibinfo{author}{\bibfnamefont{J.}~\bibnamefont{Ventura}},
  \bibinfo{author}{\bibfnamefont{J.~M.} \bibnamefont{Teixeira}},
  \bibinfo{author}{\bibfnamefont{D.~S.} \bibnamefont{Schmool}},
  \bibinfo{author}{\bibfnamefont{G.~N.} \bibnamefont{Kakazei}},
  \bibinfo{author}{\bibfnamefont{S.}~\bibnamefont{Cardoso}},
  \bibinfo{author}{\bibfnamefont{P.~P.} \bibnamefont{Freitas}},
  \bibnamefont{et~al.}, \bibinfo{journal}{Phys. Rev. B}
  \textbf{\bibinfo{volume}{91}}, \bibinfo{pages}{104407}
  (\bibinfo{year}{2015}),
  \urlprefix\url{http://link.aps.org/doi/10.1103/PhysRevB.91.104407}.

\bibitem[{\citenamefont{Averitt and Taylor}(2002)}]{averitt:1357}
\bibinfo{author}{\bibfnamefont{R.~D.} \bibnamefont{Averitt}} \bibnamefont{and}
  \bibinfo{author}{\bibfnamefont{A.~J.} \bibnamefont{Taylor}},
  \bibinfo{journal}{J. Phys.: Condens. Matter} \textbf{\bibinfo{volume}{14}},
  \bibinfo{pages}{R1357} (\bibinfo{year}{2002}).

\bibitem[{\citenamefont{Averitt et~al.}(2001)\citenamefont{Averitt, Lobad,
  Kwon, Trugman, Thorsm\o{}lle, and Taylor}}]{averitt:017401}
\bibinfo{author}{\bibfnamefont{R.~D.} \bibnamefont{Averitt}},
  \bibinfo{author}{\bibfnamefont{A.~I.} \bibnamefont{Lobad}},
  \bibinfo{author}{\bibfnamefont{C.}~\bibnamefont{Kwon}},
  \bibinfo{author}{\bibfnamefont{S.~A.} \bibnamefont{Trugman}},
  \bibinfo{author}{\bibfnamefont{V.~K.} \bibnamefont{Thorsm\o{}lle}},
  \bibnamefont{and} \bibinfo{author}{\bibfnamefont{A.~J.}
  \bibnamefont{Taylor}}, \bibinfo{journal}{Phys. Rev. Lett.}
  \textbf{\bibinfo{volume}{87}}, \bibinfo{pages}{017401}
  (\bibinfo{year}{2001}),
  \urlprefix\url{http://link.aps.org/doi/10.1103/PhysRevLett.87.017401}.

\bibitem[{\citenamefont{Silwal et~al.}(2013)\citenamefont{Silwal, La-o
  vorakiat, Chia, Kim, and Talbayev}}]{silwal:092116}
\bibinfo{author}{\bibfnamefont{P.}~\bibnamefont{Silwal}},
  \bibinfo{author}{\bibfnamefont{C.}~\bibnamefont{La-o vorakiat}},
  \bibinfo{author}{\bibfnamefont{E.~E.~M.} \bibnamefont{Chia}},
  \bibinfo{author}{\bibfnamefont{D.~H.} \bibnamefont{Kim}}, \bibnamefont{and}
  \bibinfo{author}{\bibfnamefont{D.}~\bibnamefont{Talbayev}},
  \bibinfo{journal}{AIP Advances} \textbf{\bibinfo{volume}{3}},
  \bibinfo{eid}{092116} (\bibinfo{year}{2013}),
  \urlprefix\url{http://scitation.aip.org/content/aip/journal/adva/3/9/10.1063/1.4821548}.

\bibitem[{\citenamefont{Singley et~al.}(1999)\citenamefont{Singley, Weber,
  Basov, Barry, and Coey}}]{singley:4126}
\bibinfo{author}{\bibfnamefont{E.~J.} \bibnamefont{Singley}},
  \bibinfo{author}{\bibfnamefont{C.~P.} \bibnamefont{Weber}},
  \bibinfo{author}{\bibfnamefont{D.~N.} \bibnamefont{Basov}},
  \bibinfo{author}{\bibfnamefont{A.}~\bibnamefont{Barry}}, \bibnamefont{and}
  \bibinfo{author}{\bibfnamefont{J.~M.~D.} \bibnamefont{Coey}},
  \bibinfo{journal}{Phys. Rev. B} \textbf{\bibinfo{volume}{60}},
  \bibinfo{pages}{4126} (\bibinfo{year}{1999}),
  \urlprefix\url{http://link.aps.org/doi/10.1103/PhysRevB.60.4126}.

\bibitem[{\citenamefont{Groeneveld et~al.}(1995)\citenamefont{Groeneveld,
  Sprik, and Lagendijk}}]{groeneveld:11433}
\bibinfo{author}{\bibfnamefont{R.~H.~M.} \bibnamefont{Groeneveld}},
  \bibinfo{author}{\bibfnamefont{R.}~\bibnamefont{Sprik}}, \bibnamefont{and}
  \bibinfo{author}{\bibfnamefont{A.}~\bibnamefont{Lagendijk}},
  \bibinfo{journal}{Phys. Rev. B} \textbf{\bibinfo{volume}{51}},
  \bibinfo{pages}{11433} (\bibinfo{year}{1995}),
  \urlprefix\url{http://link.aps.org/doi/10.1103/PhysRevB.51.11433}.

\bibitem[{\citenamefont{Allen}(1987)}]{allen:1460}
\bibinfo{author}{\bibfnamefont{P.~B.} \bibnamefont{Allen}},
  \bibinfo{journal}{Phys. Rev. Lett.} \textbf{\bibinfo{volume}{59}},
  \bibinfo{pages}{1460} (\bibinfo{year}{1987}),
  \urlprefix\url{http://link.aps.org/doi/10.1103/PhysRevLett.59.1460}.

\bibitem[{\citenamefont{Zhang et~al.}(2006)\citenamefont{Zhang, Nurmikko, Miao,
  Xiao, and Gupta}}]{zhang:064414}
\bibinfo{author}{\bibfnamefont{Q.}~\bibnamefont{Zhang}},
  \bibinfo{author}{\bibfnamefont{A.~V.} \bibnamefont{Nurmikko}},
  \bibinfo{author}{\bibfnamefont{G.~X.} \bibnamefont{Miao}},
  \bibinfo{author}{\bibfnamefont{G.}~\bibnamefont{Xiao}}, \bibnamefont{and}
  \bibinfo{author}{\bibfnamefont{A.}~\bibnamefont{Gupta}},
  \bibinfo{journal}{Phys. Rev. B} \textbf{\bibinfo{volume}{74}},
  \bibinfo{pages}{064414} (\bibinfo{year}{2006}),
  \urlprefix\url{http://link.aps.org/doi/10.1103/PhysRevB.74.064414}.

\bibitem[{\citenamefont{Mueller et~al.}(2009)\citenamefont{Mueller, Walowski,
  Djordjevic, Miao, Gupta, Ramos, Gehrke, Moshnyaga, Samwer, Schmalhorst
  et~al.}}]{mueller:56}
\bibinfo{author}{\bibfnamefont{G.~M.} \bibnamefont{Mueller}},
  \bibinfo{author}{\bibfnamefont{J.}~\bibnamefont{Walowski}},
  \bibinfo{author}{\bibfnamefont{M.}~\bibnamefont{Djordjevic}},
  \bibinfo{author}{\bibfnamefont{G.-X.} \bibnamefont{Miao}},
  \bibinfo{author}{\bibfnamefont{A.}~\bibnamefont{Gupta}},
  \bibinfo{author}{\bibfnamefont{A.~V.} \bibnamefont{Ramos}},
  \bibinfo{author}{\bibfnamefont{K.}~\bibnamefont{Gehrke}},
  \bibinfo{author}{\bibfnamefont{V.}~\bibnamefont{Moshnyaga}},
  \bibinfo{author}{\bibfnamefont{K.}~\bibnamefont{Samwer}},
  \bibinfo{author}{\bibfnamefont{J.}~\bibnamefont{Schmalhorst}},
  \bibnamefont{et~al.}, \bibinfo{journal}{Nat. Mater.}
  \textbf{\bibinfo{volume}{8}}, \bibinfo{pages}{56} (\bibinfo{year}{2009}).

\bibitem[{\citenamefont{Zhang et~al.}(2002)\citenamefont{Zhang, Nurmikko,
  Anguelouch, Xiao, and Gupta}}]{zhang:177402}
\bibinfo{author}{\bibfnamefont{Q.}~\bibnamefont{Zhang}},
  \bibinfo{author}{\bibfnamefont{A.~V.} \bibnamefont{Nurmikko}},
  \bibinfo{author}{\bibfnamefont{A.}~\bibnamefont{Anguelouch}},
  \bibinfo{author}{\bibfnamefont{G.}~\bibnamefont{Xiao}}, \bibnamefont{and}
  \bibinfo{author}{\bibfnamefont{A.}~\bibnamefont{Gupta}},
  \bibinfo{journal}{Phys. Rev. Lett.} \textbf{\bibinfo{volume}{89}},
  \bibinfo{pages}{177402} (\bibinfo{year}{2002}),
  \urlprefix\url{http://link.aps.org/doi/10.1103/PhysRevLett.89.177402}.

\bibitem[{\citenamefont{Carpene et~al.}(2013)\citenamefont{Carpene, Boschini,
  Hedayat, Piovera, Dallera, Puppin, Mansurova, M\"unzenberg, Zhang, and
  Gupta}}]{carpene:174437}
\bibinfo{author}{\bibfnamefont{E.}~\bibnamefont{Carpene}},
  \bibinfo{author}{\bibfnamefont{F.}~\bibnamefont{Boschini}},
  \bibinfo{author}{\bibfnamefont{H.}~\bibnamefont{Hedayat}},
  \bibinfo{author}{\bibfnamefont{C.}~\bibnamefont{Piovera}},
  \bibinfo{author}{\bibfnamefont{C.}~\bibnamefont{Dallera}},
  \bibinfo{author}{\bibfnamefont{E.}~\bibnamefont{Puppin}},
  \bibinfo{author}{\bibfnamefont{M.}~\bibnamefont{Mansurova}},
  \bibinfo{author}{\bibfnamefont{M.}~\bibnamefont{M\"unzenberg}},
  \bibinfo{author}{\bibfnamefont{X.}~\bibnamefont{Zhang}}, \bibnamefont{and}
  \bibinfo{author}{\bibfnamefont{A.}~\bibnamefont{Gupta}},
  \bibinfo{journal}{Phys. Rev. B} \textbf{\bibinfo{volume}{87}},
  \bibinfo{pages}{174437} (\bibinfo{year}{2013}),
  \urlprefix\url{http://link.aps.org/doi/10.1103/PhysRevB.87.174437}.

\bibitem[{\citenamefont{M\"uller et~al.}(2008)\citenamefont{M\"uller,
  M\"unzenberg, Miao, and Gupta}}]{munzenberg:020412}
\bibinfo{author}{\bibfnamefont{G.~M.} \bibnamefont{M\"uller}},
  \bibinfo{author}{\bibfnamefont{M.}~\bibnamefont{M\"unzenberg}},
  \bibinfo{author}{\bibfnamefont{G.-X.} \bibnamefont{Miao}}, \bibnamefont{and}
  \bibinfo{author}{\bibfnamefont{A.}~\bibnamefont{Gupta}},
  \bibinfo{journal}{Phys. Rev. B} \textbf{\bibinfo{volume}{77}},
  \bibinfo{pages}{020412} (\bibinfo{year}{2008}),
  \urlprefix\url{http://link.aps.org/doi/10.1103/PhysRevB.77.020412}.

\bibitem[{\citenamefont{Zhao et~al.}(2011)\citenamefont{Zhao, Talbayev, Ma,
  Ren, Venimadhav, Li, and L\"upke}}]{zhao:207205}
\bibinfo{author}{\bibfnamefont{H.~B.} \bibnamefont{Zhao}},
  \bibinfo{author}{\bibfnamefont{D.}~\bibnamefont{Talbayev}},
  \bibinfo{author}{\bibfnamefont{X.}~\bibnamefont{Ma}},
  \bibinfo{author}{\bibfnamefont{Y.~H.} \bibnamefont{Ren}},
  \bibinfo{author}{\bibfnamefont{A.}~\bibnamefont{Venimadhav}},
  \bibinfo{author}{\bibfnamefont{Q.}~\bibnamefont{Li}}, \bibnamefont{and}
  \bibinfo{author}{\bibfnamefont{G.}~\bibnamefont{L\"upke}},
  \bibinfo{journal}{Phys. Rev. Lett.} \textbf{\bibinfo{volume}{107}},
  \bibinfo{pages}{207205} (\bibinfo{year}{2011}),
  \urlprefix\url{http://link.aps.org/doi/10.1103/PhysRevLett.107.207205}.

\bibitem[{\citenamefont{Talbayev et~al.}(2006)\citenamefont{Talbayev, Zhao,
  L\"upke, Venimadhav, and Li}}]{talbayev:014417}
\bibinfo{author}{\bibfnamefont{D.}~\bibnamefont{Talbayev}},
  \bibinfo{author}{\bibfnamefont{H.}~\bibnamefont{Zhao}},
  \bibinfo{author}{\bibfnamefont{G.}~\bibnamefont{L\"upke}},
  \bibinfo{author}{\bibfnamefont{A.}~\bibnamefont{Venimadhav}},
  \bibnamefont{and} \bibinfo{author}{\bibfnamefont{Q.}~\bibnamefont{Li}},
  \bibinfo{journal}{Phys. Rev. B} \textbf{\bibinfo{volume}{73}},
  \bibinfo{pages}{014417} (\bibinfo{year}{2006}),
  \urlprefix\url{http://link.aps.org/doi/10.1103/PhysRevB.73.014417}.

\bibitem[{\citenamefont{Talbayev et~al.}(2005)\citenamefont{Talbayev, Zhao,
  L\"upke, Chen, and Li}}]{talbayev:182501}
\bibinfo{author}{\bibfnamefont{D.}~\bibnamefont{Talbayev}},
  \bibinfo{author}{\bibfnamefont{H.}~\bibnamefont{Zhao}},
  \bibinfo{author}{\bibfnamefont{G.}~\bibnamefont{L\"upke}},
  \bibinfo{author}{\bibfnamefont{J.}~\bibnamefont{Chen}}, \bibnamefont{and}
  \bibinfo{author}{\bibfnamefont{Q.}~\bibnamefont{Li}}, \bibinfo{journal}{Appl.
  Phys. Lett.} \textbf{\bibinfo{volume}{86}}, \bibinfo{eid}{182501}
  (\bibinfo{year}{2005}),
  \urlprefix\url{http://scitation.aip.org/content/aip/journal/apl/86/18/10.1063/1.1922073}.

\bibitem[{\citenamefont{Stewart et~al.}(2009)\citenamefont{Stewart, Chetry,
  Chapler, Qazilbash, Schafgans, Gupta, Tiwald, and Basov}}]{stewart:144414}
\bibinfo{author}{\bibfnamefont{M.~K.} \bibnamefont{Stewart}},
  \bibinfo{author}{\bibfnamefont{K.~B.} \bibnamefont{Chetry}},
  \bibinfo{author}{\bibfnamefont{B.}~\bibnamefont{Chapler}},
  \bibinfo{author}{\bibfnamefont{M.~M.} \bibnamefont{Qazilbash}},
  \bibinfo{author}{\bibfnamefont{A.~A.} \bibnamefont{Schafgans}},
  \bibinfo{author}{\bibfnamefont{A.}~\bibnamefont{Gupta}},
  \bibinfo{author}{\bibfnamefont{T.~E.} \bibnamefont{Tiwald}},
  \bibnamefont{and} \bibinfo{author}{\bibfnamefont{D.~N.} \bibnamefont{Basov}},
  \bibinfo{journal}{Phys. Rev. B} \textbf{\bibinfo{volume}{79}},
  \bibinfo{pages}{144414} (\bibinfo{year}{2009}),
  \urlprefix\url{http://link.aps.org/doi/10.1103/PhysRevB.79.144414}.

\bibitem[{\citenamefont{Iliev et~al.}(1999)\citenamefont{Iliev, Litvinchuk,
  Lee, Chu, Barry, and Coey}}]{iliev:33}
\bibinfo{author}{\bibfnamefont{M.~N.} \bibnamefont{Iliev}},
  \bibinfo{author}{\bibfnamefont{A.~P.} \bibnamefont{Litvinchuk}},
  \bibinfo{author}{\bibfnamefont{H.-G.} \bibnamefont{Lee}},
  \bibinfo{author}{\bibfnamefont{C.~W.} \bibnamefont{Chu}},
  \bibinfo{author}{\bibfnamefont{A.}~\bibnamefont{Barry}}, \bibnamefont{and}
  \bibinfo{author}{\bibfnamefont{J.~M.~D.} \bibnamefont{Coey}},
  \bibinfo{journal}{Phys. Rev. B} \textbf{\bibinfo{volume}{60}},
  \bibinfo{pages}{33} (\bibinfo{year}{1999}),
  \urlprefix\url{http://link.aps.org/doi/10.1103/PhysRevB.60.33}.

\bibitem[{\citenamefont{Li et~al.}(1999)\citenamefont{Li, Gupta, McGuire,
  Duncombe, and Xiao}}]{li:5585}
\bibinfo{author}{\bibfnamefont{X.~W.} \bibnamefont{Li}},
  \bibinfo{author}{\bibfnamefont{A.}~\bibnamefont{Gupta}},
  \bibinfo{author}{\bibfnamefont{T.~R.} \bibnamefont{McGuire}},
  \bibinfo{author}{\bibfnamefont{P.~R.} \bibnamefont{Duncombe}},
  \bibnamefont{and} \bibinfo{author}{\bibfnamefont{G.}~\bibnamefont{Xiao}},
  \bibinfo{journal}{Journal of Applied Physics} \textbf{\bibinfo{volume}{85}}
  (\bibinfo{year}{1999}).

\bibitem[{\citenamefont{Schiffer et~al.}(1995)\citenamefont{Schiffer, Ramirez,
  Bao, and Cheong}}]{schiffer:3336}
\bibinfo{author}{\bibfnamefont{P.}~\bibnamefont{Schiffer}},
  \bibinfo{author}{\bibfnamefont{A.~P.} \bibnamefont{Ramirez}},
  \bibinfo{author}{\bibfnamefont{W.}~\bibnamefont{Bao}}, \bibnamefont{and}
  \bibinfo{author}{\bibfnamefont{S.-W.} \bibnamefont{Cheong}},
  \bibinfo{journal}{Phys. Rev. Lett.} \textbf{\bibinfo{volume}{75}},
  \bibinfo{pages}{3336} (\bibinfo{year}{1995}),
  \urlprefix\url{http://link.aps.org/doi/10.1103/PhysRevLett.75.3336}.

\bibitem[{\citenamefont{Okimoto et~al.}(1995)\citenamefont{Okimoto, Katsufuji,
  Ishikawa, Urushibara, Arima, and Tokura}}]{okimoto:109}
\bibinfo{author}{\bibfnamefont{Y.}~\bibnamefont{Okimoto}},
  \bibinfo{author}{\bibfnamefont{T.}~\bibnamefont{Katsufuji}},
  \bibinfo{author}{\bibfnamefont{T.}~\bibnamefont{Ishikawa}},
  \bibinfo{author}{\bibfnamefont{A.}~\bibnamefont{Urushibara}},
  \bibinfo{author}{\bibfnamefont{T.}~\bibnamefont{Arima}}, \bibnamefont{and}
  \bibinfo{author}{\bibfnamefont{Y.}~\bibnamefont{Tokura}},
  \bibinfo{journal}{Phys. Rev. Lett.} \textbf{\bibinfo{volume}{75}},
  \bibinfo{pages}{109} (\bibinfo{year}{1995}),
  \urlprefix\url{http://link.aps.org/doi/10.1103/PhysRevLett.75.109}.

\bibitem[{\citenamefont{Kim et~al.}(1998)\citenamefont{Kim, Jung, and
  Noh}}]{kim:1517}
\bibinfo{author}{\bibfnamefont{K.~H.} \bibnamefont{Kim}},
  \bibinfo{author}{\bibfnamefont{J.~H.} \bibnamefont{Jung}}, \bibnamefont{and}
  \bibinfo{author}{\bibfnamefont{T.~W.} \bibnamefont{Noh}},
  \bibinfo{journal}{Phys. Rev. Lett.} \textbf{\bibinfo{volume}{81}},
  \bibinfo{pages}{1517} (\bibinfo{year}{1998}),
  \urlprefix\url{http://link.aps.org/doi/10.1103/PhysRevLett.81.1517}.

\bibitem[{\citenamefont{Lee et~al.}(1999)\citenamefont{Lee, Jung, Lee, Ahn,
  Noh, Kim, and Cheong}}]{lee:5251}
\bibinfo{author}{\bibfnamefont{H.~J.} \bibnamefont{Lee}},
  \bibinfo{author}{\bibfnamefont{J.~H.} \bibnamefont{Jung}},
  \bibinfo{author}{\bibfnamefont{Y.~S.} \bibnamefont{Lee}},
  \bibinfo{author}{\bibfnamefont{J.~S.} \bibnamefont{Ahn}},
  \bibinfo{author}{\bibfnamefont{T.~W.} \bibnamefont{Noh}},
  \bibinfo{author}{\bibfnamefont{K.~H.} \bibnamefont{Kim}}, \bibnamefont{and}
  \bibinfo{author}{\bibfnamefont{S.-W.} \bibnamefont{Cheong}},
  \bibinfo{journal}{Phys. Rev. B} \textbf{\bibinfo{volume}{60}},
  \bibinfo{pages}{5251} (\bibinfo{year}{1999}),
  \urlprefix\url{http://link.aps.org/doi/10.1103/PhysRevB.60.5251}.

\end{thebibliography}

\end{document}